\documentclass[
    a4paper,
    onecolumn,
    11pt,
    accepted=2022-07-17,
    amsmath
]{quantumarticle}
\pdfoutput=1
\usepackage[utf8]{inputenc}
\usepackage[T1]{fontenc}
\usepackage[english]{babel}
\usepackage{booktabs}
\usepackage{microtype}
\usepackage{xcolor}
\usepackage{hyperref}
\usepackage{cleveref}
\crefname{lstlisting}{codeblock}{codeblock}
\Crefname{lstlisting}{Codeblock}{Codeblock}

\usepackage{listings}

\definecolor{codegreen}{rgb}{0,0.6,0}
\definecolor{codegray}{rgb}{0.5,0.5,0.5}
\definecolor{codepurple}{rgb}{0.58,0,0.82}
\definecolor{backcolour}{rgb}{0.95,0.95,0.92}

\lstdefinestyle{mystyle}{
    backgroundcolor=\color{lightgray!30},   
    commentstyle=\color{codegreen},
    keywordstyle=\color{magenta},
    numberstyle=\tiny\color{codegray},
    stringstyle=\color{codepurple},
    basicstyle=\footnotesize\ttfamily,
    breakatwhitespace=false,         
    breaklines=true,                 
    captionpos=b,                    
    keepspaces=true,                 
    numbers=left,                    
    numbersep=5pt,                  
    showspaces=false,                
    showstringspaces=false,
    showtabs=false,                  
    tabsize=2,
}
\lstdefinestyle{nonumbers}{
    backgroundcolor=\color{lightgray!30},   
    commentstyle=\color{codegreen},
    keywordstyle=\color{magenta},
    numberstyle=\tiny\color{codegray},
    stringstyle=\color{codepurple},
    basicstyle=\footnotesize\ttfamily,
    breakatwhitespace=false,         
    breaklines=true,                 
    captionpos=b,                    
    keepspaces=true,                 
    numbers=none,                    
    numbersep=5pt,                  
    showspaces=false,                
    showstringspaces=false,
    showtabs=false,                  
    tabsize=2,
}

\newcommand{\inlinecode}[1]{\texttt{#1}}

\begin{document}

\title{Mitiq: A software package for error mitigation on noisy quantum computers}

\author{Ryan LaRose}
\affiliation{Unitary Fund}
\affiliation{Michigan State University, East Lansing, MI}

\author{Andrea Mari}
\affiliation{Unitary Fund}

\author{Sarah Kaiser}
\affiliation{Unitary Fund}

\author{Peter J. Karalekas}
\altaffiliation[Current address:]{ AWS Center for Quantum Computing, Pasadena, CA 91125, USA}
\affiliation{Unitary Fund}

\author{Andre A. Alves}
\affiliation{Hamburg University of Applied Sciences, Hamburg, Germany}

\author{Piotr Czarnik}
\affiliation{Theoretical Division, Los Alamos National Laboratory, Los Alamos, NM 87545, USA}

\author{Mohamed El Mandouh}
\affiliation{Institute for Quantum Computing, University of Waterloo, Waterloo, ON, N2L 3G1, Canada}

\author{Max H. Gordon}
\affiliation{Instituto de Física Teórica, UAM/CSIC, Universidad Autónoma de Madrid, Madrid, Spain}

\author{Yousef Hindy}
\affiliation{Stanford University, Palo Alto, CA}

\author{Aaron Robertson}
\affiliation{Independent researcher}

\author{Purva Thakre}
\affiliation{Southern Illinois University, Carbondale, IL}


\author{Misty Wahl}
\affiliation{Unitary Fund}

\author{Danny Samuel}
\affiliation{Unitary Fund}

\author{Rahul Mistri}
\affiliation{Unitary Fund}

\author{Maxime Tremblay}
\affiliation{Institut quantique, Université de Sherbrooke, Sherbrooke, QC, J1K 2R1, Canada}

\author{Nick Gardner}
\affiliation{Stanford University, Palo Alto, CA}

\author{Nathaniel T. Stemen}
\affiliation{Unitary Fund}


\author{Nathan Shammah}
\affiliation{Unitary Fund}

\author{William J. Zeng}
\affiliation{Unitary Fund}
\affiliation{Stanford University, Palo Alto, CA}
\affiliation{Goldman, Sachs \& Co, New York, NY}

\maketitle

\begin{abstract}
    We introduce Mitiq, a Python package for error mitigation on noisy quantum computers.
    Error mitigation techniques can reduce the impact of noise on near-term quantum computers with minimal overhead in quantum resources by relying on a mixture
    of quantum sampling and classical post-processing techniques. Mitiq is an extensible toolkit of different error mitigation methods, including  zero-noise
    extrapolation, probabilistic error cancellation, and Clifford data regression. The library is designed to be compatible with generic backends and
    interfaces with different quantum software frameworks.
    We describe Mitiq using code snippets to demonstrate usage and discuss features and contribution guidelines.
    We present several examples demonstrating error mitigation on IBM and Rigetti superconducting quantum processors as well as on noisy simulators.
\end{abstract}

\section{Introduction}

Methods to counteract noise are critical for realizing practical quantum computation. While fault-tolerant quantum computers that use error-correcting codes are an ideal goal, they require physical resources beyond current experimental capabilities. It is therefore interesting and important to develop other methods for dealing with noise on near-term quantum computers.

In recent years, several methods, collectively referred to as quantum error mitigation methods~\cite{Endo_2021}, have been proposed and developed for this task. Among them are zero-noise extrapolation~\cite{Temme_2017_PRL,Li_2017_PRX}, probabilistic error cancellation~\cite{Temme_2017_PRL, endo2018practical}, Clifford data regression~\cite{czarnik2020error, lowe2020unified}, dynamical decoupling~\cite{santos2005dynamical, viola2005random, pokharel2018demonstration}, randomized compiling~\cite{wallman2016noise}, and subspace expansion~\cite{McClean_2020_NatComm}.
Several error mitigation methods have also been tested experimentally~\cite{Kandala_2019_Nature,Giurgica_2020_arxiv,Urbanek_Nachman_de_Jong_2019,Vuillot,Arute_2020_Science,Song_2019_Science_Advances,Zhang_2020_NatComm}. To aid research, improve reproducibility, and move towards practical applications, it is important to have a unified framework for implementing error mitigation techniques on multiple quantum back-ends.

\begin{figure}
    \centering
    \includegraphics[width=0.6\textwidth]{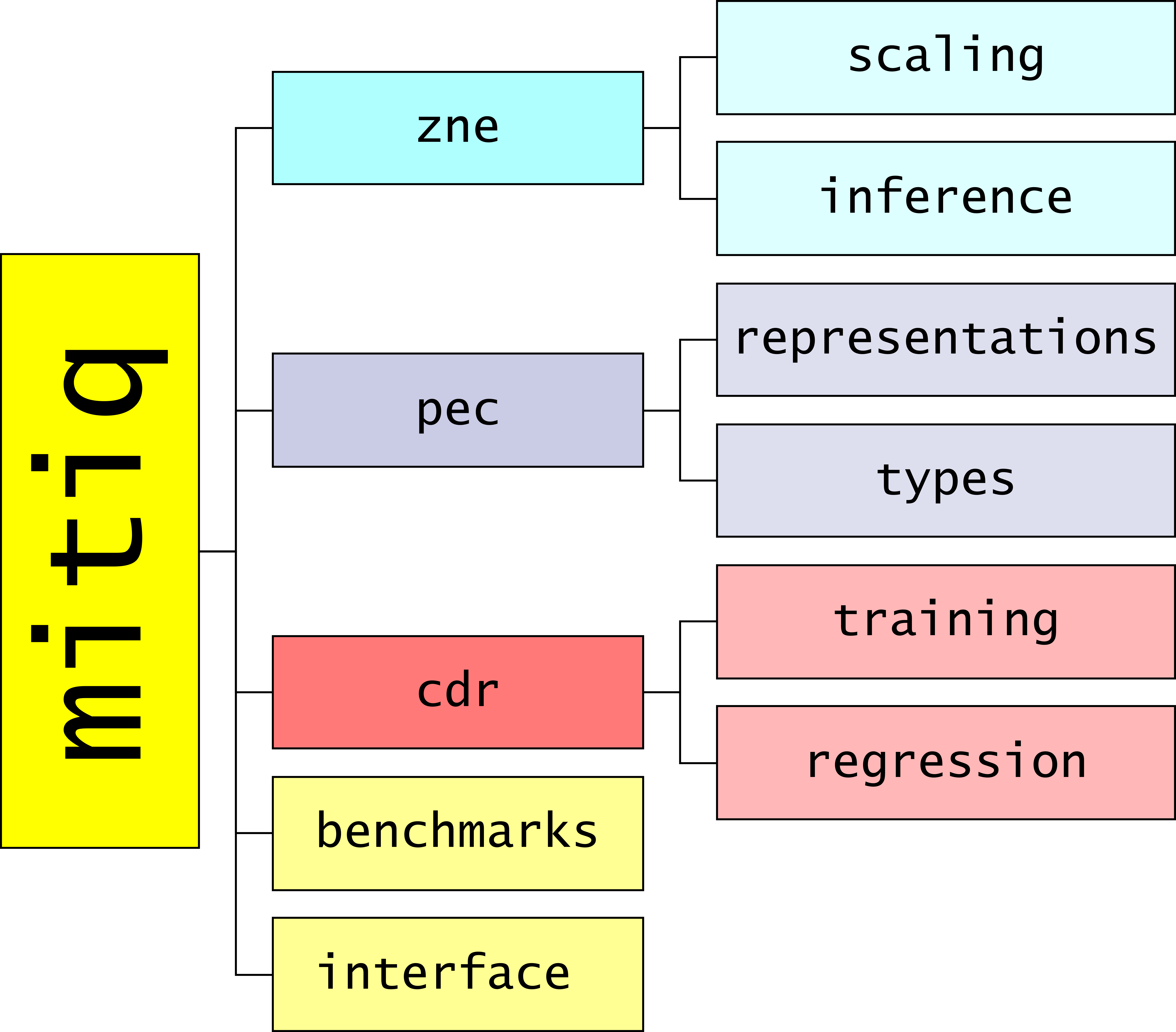}
    \caption{The structure of Mitiq modules. Different error mitigation techniques are organized in different modules, including zero noise extrapolation (\inlinecode{zne}), probabilistic error cancellation (\inlinecode{pec}), and Clifford data regression (\inlinecode{cdr}). Other modules are dedicated to auxiliary tasks such as interfacing with different quantum software libraries (\inlinecode{interface}) and bench-marking error mitigation strategies (\inlinecode{benchmarks}).}\label{fig:mitiq-tree}
\end{figure}

To these ends, we introduce Mitiq: a software package for error mitigation on noisy quantum computers. Mitiq is an open-source Python library that interfaces with multiple quantum programming front-ends to implement error mitigation techniques on various real and simulated quantum processors. Mitiq supports Cirq~\cite{cirq_blog}, Qiskit~\cite{Qiskit}, pyQuil~\cite{Smith_2016_arxiv}, and Braket~\cite{Braket} circuit types and any  back-ends, real or simulated, that can execute them. The library is extensible in that new front-ends and back-ends can be easily supported as they become available. Mitiq currently implements zero-noise extrapolation (ZNE), probabilistic error cancellation (PEC), and Clifford data regression (CDR), and its modular design allows support for additional techniques, as shown in \cref{fig:mitiq-tree}. Error mitigation methods can be applied in a few additional lines of code, but the library is still flexible enough for advanced usage.

In \cref{sec:getting-started}, we show how to get started with Mitiq and illustrate its main usage. We then show experimental and numerical examples in \cref{sec:examples-and-benchmarks} that demonstrate how error mitigation with Mitiq improves the performance of noisy quantum computations. In \cref{sec:zne}, we describe in detail the zero-noise extrapolation module. In \cref{sec:pec}, we give an overview of the probabilistic error cancellation module.
In \cref{sec:cdr}, we present the Clifford data regression module.
We discuss further software details and library information in \cref{sec:software-details-library-information} including future development, contribution guidelines, and planned maintenance and support. Finally, in \cref{sec:discussion} we discuss the relationship between Mitiq and other techniques for dealing with errors in quantum computers.

\begin{table}[t]
    \centering
    \begin{tabular}{c c}
        \textbf{Software Framework} & \textbf{Circuit Type}            \\
        \toprule
        Cirq                        & \texttt{cirq.Circuit}            \\
        Qiskit                      & \texttt{qiskit.QuantumCircuit}   \\
        PyQuil                      & \texttt{pyquil.Program}          \\
        Braket                      & \texttt{braket.circuits.Circuit}
    \end{tabular}
    \caption{The quantum software frameworks compatible with Mitiq. Since Mitiq interacts with circuits but is not directly responsible for their execution, supporting a new circuit type requires only to define a few conversion functions. Therefore, we expect the list in this table to grow in the future.}\label{tab:backends}
\end{table}

\section{Getting started with Mitiq}\label{sec:getting-started}

\subsection{Requirements and installation}\label{sec:requirements-and-installation}

Mitiq is a Python library that can be installed on Mac, Windows, and Linux operating systems via PyPI by executing the instruction below at a command line.
\begin{lstlisting}[language=bash, caption={Installing Mitiq through PyPI.}, firstnumber=1]
pip install mitiq
\end{lstlisting}
To test installation, one can run the following.
\begin{lstlisting}[language=Python, caption={Testing installation \& viewing package versions.}, firstnumber=1, label=lst:mitiq.about]
import mitiq

mitiq.about()
\end{lstlisting}
This code prints information about the Mitiq version and the versions of installed packages.
\begin{lstlisting}[caption={Example output of \cref{lst:mitiq.about}.}, style=nonumbers, basicstyle=\ttfamily\scriptsize]
Mitiq: A Python toolkit for implementing error mitigation on quantum computers
==============================================================================
Authored by: Mitiq team, 2020 & later (https://github.com/unitaryfund/mitiq)

Mitiq Version:  0.9.3

Core Dependencies
-----------------
Cirq Version:   0.10.0
NumPy Version:  1.20.1
SciPy Version:  1.4.1

Optional Dependencies
---------------------
PyQuil Version: 2.28.0
Qiskit Version: 0.24.0
Braket Version: 1.5.16

Python Version: 3.7.7
Platform Info:  Linux (x86_64)
\end{lstlisting}
%

In this example output, we see several packages. The core requirements of Mitiq are Cirq (used to internally represent and manipulate quantum circuits), NumPy (used for general numerical procedures), and SciPy~\cite{Virtanen_2020_Nature_Methods} (used for curve fitting). The remaining packages (pyQuil, Qiskit, Braket) are optional quantum software packages which can interface with Mitiq.
Although Mitiq's internal quantum circuit representation is a Cirq \inlinecode{Circuit}, any supported quantum circuit types can be used with Mitiq. The current supported circuit types are summarized in \cref{tab:backends}. A Mitiq \inlinecode{QPROGRAM} is the union of all supported circuit representations which are installed with Mitiq. For example, if  Qiskit is the only optional package installed, the \inlinecode{QPROGRAM} type will be the union of a Cirq \inlinecode{Circuit} and a Qiskit \inlinecode{QuantumCircuit}. If pyQuil is also installed, \inlinecode{QPROGRAM} will also include the pyQuil \inlinecode{Program} type.
%

The source code for Mitiq is hosted on GitHub at
\begin{center}
    \url{https://github.com/unitaryfund/mitiq}
\end{center}
and is distributed with an open-source software license: GNU GPL v.\ 3.0.

More details about the software, packaging information, and guidelines for contributing to Mitiq are included in \cref{sec:software-details-library-information}.

\subsection{Main usage}\label{sec:main-usage}

To implement error mitigation techniques in Mitiq, we assume that the user has a function which inputs a quantum circuit and returns the expectation value of an observable. Mitiq uses this function as an abstract interface of a generic noisy backend and we refer to it as an \textit{executor} because it \textit{executes} a quantum circuit. The signature of this function should be as follows:
\begin{lstlisting}[language=python, caption={Signature of an executor function which is used by Mitiq to perform quantum error mitigation.},
label={lst:executor_signature},
firstnumber=1]
def executor(circuit: mitiq.QPROGRAM) -> float:
\end{lstlisting}
%

Mitiq treats the \inlinecode{executor} as a black box to mitigate the expectation value of the observable returned by this function. The user is responsible for defining the body of the \inlinecode{executor}, which generally involves:
\begin{enumerate}
    \item Running the \inlinecode{circuit} on a real or simulated QPU.
    \item Post-processing to compute an expectation value.
    \item Returning the expectation value as a floating-point number.
\end{enumerate}
Example \inlinecode{executor} functions are shown in \cref{sec:executors}.
Since Mitiq treats the \inlinecode{executor} as a black box, circuits can be run on any quantum processor available to the user. For example, we present benchmarks run on IBM and Rigetti quantum processors as well as on noisy simulators in \cref{sec:examples-and-benchmarks}.

\begin{figure}
    \centering
    \includegraphics[width=0.75\textwidth]{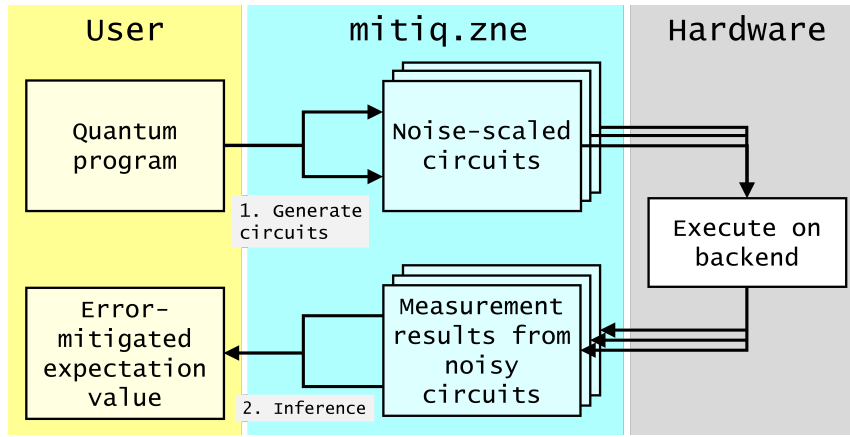}
    \caption{Overview of the zero-noise extrapolation workflow in Mitiq. An input quantum program is converted into a set of noise-scaled circuits defined by a noise scaling method and a set of noise scale factors. These auxiliary circuits are executed on the back-end according to a user-defined executor function (see \cref{sec:executors} for examples) producing set of noise-scaled expectation values. A classical inference technique is used to fit a model to these noise-scaled expectation values. Once the best-fit model is established, the zero-noise limit is returned to give an error-mitigated expectation value.}\label{fig:zne-pipeline}
\end{figure}

Once the executor is defined, implementing a standard error mitigation technique such as zero-noise extrapolation (ZNE) needs only a single line of code:
\begin{lstlisting}[language=Python, caption={Using Mitiq to perform zero-noise extrapolation. The \inlinecode{circuit} is a supported quantum program type, and the \inlinecode{executor} is a function which executes the circuit and returns an expectation value.}, firstnumber=1]
from mitiq.zne import execute_with_zne

zne_value = execute_with_zne(circuit, executor)
\end{lstlisting}
The \inlinecode{execute\_with\_zne} function uses the \inlinecode{executor} to evaluate the input \inlinecode{circuit} at different noise levels, extrapolates back to the zero-noise limit and then returns this value as an estimate of the noiseless observable.
\Cref{fig:zne-pipeline} shows a high-level workflow.

As described in \cref{sec:zne}, there are multiple techniques to scale the noise in a quantum circuit and infer (extrapolate back to) the zero-noise limit. The default noise scaling method used by \inlinecode{execute\_with\_zne} is random local unitary folding~\cite{Giurgica_2020_arxiv} (see \cref{sec:noise-scaling}), and the default inference technique is Richardson extrapolation (see \cref{sec:factories}). Different techniques can be specified as arguments to \inlinecode{execute\_with\_zne} as follows.

\begin{lstlisting}[
    language=Python,
    caption={Providing arguments to \inlinecode{execute\_with\_zne} to use different noise scaling methods and inference techniques.},
    firstnumber=1]
zne_value = execute_with_zne(
    circuit,
    executor,
    scale_noise=<noise scaling method>,
    factory=<inference method>,
)
\end{lstlisting}

In addition to zero-noise extrapolation, one might be interested in applying a different error mitigation technique.
For example, probabilistic error cancellation (PEC)~\cite{Temme_2017_PRL, endo2018practical} is a method which promises to reduce the noise of a quantum computer with the only additional resource requirement being a higher sampling overhead.

Assuming the user has defined an \inlinecode{executor} as described above, PEC can be applied as follows:

\begin{lstlisting}[
    language=Python,
    caption={Using Mitiq to perform probabilistic error cancellation. The \inlinecode{circuit} is a supported quantum program type, the \inlinecode{executor} is a function which executes the circuit and returns an expectation value and the \inlinecode{representations} argument contains information about the quasi-probability representations of the ideal gates in terms of the hardware noisy gates. This Codeblock is a template --- a complete, executable example can be found in the Mitiq documentation (see \cref{sec:documentation}).},
    label=lst:execute_with_pec]
from mitiq.pec import execute_with_pec

pec_value = execute_with_pec(
    circuit,
    executor,
    representations=<quasi-probability representations of ideal circuit gates>,
)
\end{lstlisting}

The \inlinecode{execute\_with\_pec} function internally samples from a quasi-probability representation of the input \inlinecode{circuit} that depends on the input \inlinecode{representations} of individual gates (see \cref{sec:representations} for more details on gate representations).
The user-defined \inlinecode{executor} is used to run the sampled circuits.
Eventually, \inlinecode{execute\_with\_pec} combines the results and returns an unbiased estimate of the ideal observable. As schematically represented in \cref{fig:pec-pipeline}, the workflow is very similar to the previous case of ZNE (shown in \cref{fig:zne-pipeline}) but, in this case, the noisy circuits are sampled probabilistically and executed at the base noise level of the underlying hardware (noise scaling is not used).

The code examples shown in \crefrange{lst:executor_signature}{lst:execute_with_pec} demonstrate the main usage of Mitiq. Alternatives to the \inlinecode{execute\_with\_zne} and \inlinecode{execute\_with\_pec} functions are described in \cref{sec:alternate-ways-to-use-zne} --- these alternatives implement the same methods but offer different ways to call them which may be more convenient, depending on context.

In the following section, we show results of benchmarks using Mitiq on IBM and Rigetti quantum processors as well as noisy simulators. We then explain the structure of the library in more detail.

\section{Benchmarks with Mitiq}\label{sec:examples-and-benchmarks}

\subsection{Randomized benchmarking circuits}

\begin{figure}[ht]
    \centering
    \includegraphics[width=0.9\textwidth]{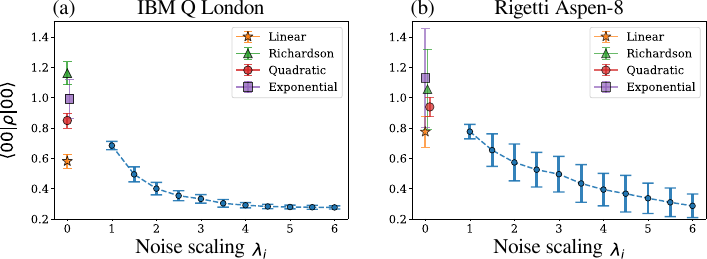}
    \caption{Zero-noise extrapolation on two-qubit randomized benchmarking circuits run on (a) the IBMQ ``London'' quantum processor and (b) the Rigetti Aspen-8 quantum processor. Results are obtained from 50 randomized benchmarking circuits which contain, on average, 97 single-qubit gates and 17 two-qubit gates for (a) and 19 single-qubit gates and 7 two-qubit gates for (b). Noise is increased via random local unitary folding (see \cref{sec:noise-scaling}), and markers show zero-noise values obtained by different extrapolation techniques (see \cref{sec:factories}). For example, the red circle is obtained by fitting a quadratic polynomial to the data points (blue), whereas the purple square is obtained by fitting an exponential decay to the same data points. (Note that some markers are staggered for visualization, but all are extrapolated to the zero-noise limit.) In this example, the true zero-noise value is $\langle 0 0 | \rho | 0 0 \rangle = 1$. For (b), qubits 32 and 33 are used on the Aspen-8 processor, while for (a) the same two qubits are not necessarily used for each run.  For linear, quadratic and exponential extrapolations, all data points are used to fit the corresponding extrapolation functions. For Richardson extrapolation, we use only three data points (first, middle, and last), corresponding to a quadratic interpolation of the three points.}\label{fig:ibm-rigetti-randomized-benchmark}
\end{figure}

\Cref{fig:ibm-rigetti-randomized-benchmark} shows the effect of zero-noise extrapolation on two-qubit randomized benchmarking circuits run on both IBM and Rigetti quantum computers. The blue curve shows the expectation value $\langle 0 0 | \rho | 0 0 \rangle$ (which should be 1 for a noiseless circuit where $\rho = |0 0\rangle \langle 0 0|$) at different noise levels, and markers show mitigated observable values obtained from different inference techniques. Error bars show the standard deviation over fifty independent runs.

Depending on the noise model as well as base noise level, different inference techniques can provide better zero-noise estimates.
The aim of the experiments shown in \cref{fig:ibm-rigetti-randomized-benchmark} is to demonstrate how Mitiq can be used to easily apply different extrapolation techniques on different backends. In this work,  we are not interested in a rigorous comparison of the performances of different extrapolation methods, since this would require a much more detailed experimental and statistical analysis.

We discuss inference techniques more in \cref{sec:factories} and the limitations of zero-noise extrapolation more in \cref{subsec:zne-limitations}.

\subsection{Potential energy surface of \texorpdfstring{$H_2$}{H\_2}}

We now consider a canonical example of computing the potential energy surface of molecular Hydrogen using the variational quantum eigensolver. We follow Ref.~\cite{scalable_2016} and use the minimal STO-6G basis and Bravyi-Kitaev transformation to write the Hamiltonian for $\text{H}_2$ as
\begin{equation}\label{eqn:h2-hamiltonian}
    H = g_0 I + g_1 Z_0 + g_2 Z_1 + g_3 Z_0 Z_1 + g_4 X_0 X_1 + g_5 Y_0 Y_1 .
\end{equation}
Here, $g_i$ are numerical coefficients which depend on the atomic separation and $I, X, Y$ and $Z$ are Pauli operators.
We use the same single-parameter variational circuit shown in fig.~1 of Ref.~\cite{scalable_2016} and we minimize the expectation value of the  Hamiltonian given in  Eq.~\eqref{eqn:h2-hamiltonian} via independent brute force optimizations evaluated for different values of the atomic distance (bond length). We simulate the experiment with and without error mitigation, assuming the presence of single-qubit depolarizing noise with error probability $p = 0.05$ (acting after each layer of gates).

\begin{figure}[ht]
    \centering
    \includegraphics[width=0.9\textwidth]{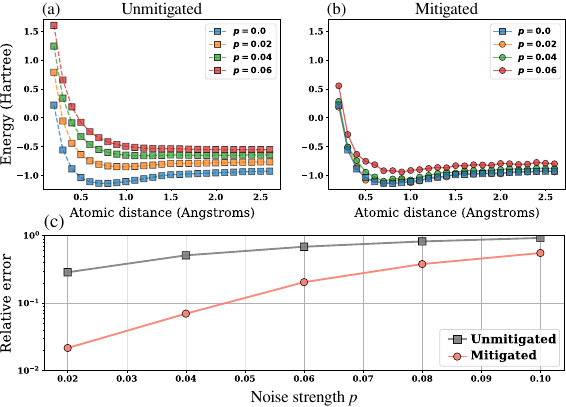}
    \caption{Unmitigated (a) and mitigated (b) energy surfaces of $H_2$. The mitigated energy surfaces use zero-noise extrapolation with random local unitary folding (see \cref{sec:noise-scaling}) and second-order polynomial inference (see \cref{sec:factories}). Panel (c) quantifies the relative error of potential energy surfaces as the $L_2$ distance $\| E_0(r) - E_p(r) \|_2 / \| E_0 (r) \|_2$ for different (simulated) depolarizing noise strengths $p$. In the previous formula, $E_0$ is the ideal noiseless ($p = 0$) expectation value of the Hamiltonian, and $r$ denotes the inter-atomic distance. The code to reproduce the results of this Figure can be found in the \textit{Examples} section of the Mitiq documentation.}\label{fig:h2-energy-surface}
\end{figure}

\Cref{fig:h2-energy-surface}(a) shows unmitigated energy surfaces at three different noise levels while \cref{fig:h2-energy-surface}(b) shows the mitigated energy surfaces. To compute the mitigated curves, we use zero-noise extrapolation with random local unitary folding (see \cref{sec:noise-scaling}) and second-order polynomial inference (see \cref{sec:factories}). As can be seen, the mitigated curves overlap with the true noiseless curve much more closely than the unmitigated curves. The error is quantified in \cref{fig:h2-energy-surface}(c).

\subsection{Probabilistic error cancellation example}\label{sec:pec-example}

\begin{figure}[t]
    \centering
    \includegraphics[width=0.7\textwidth]{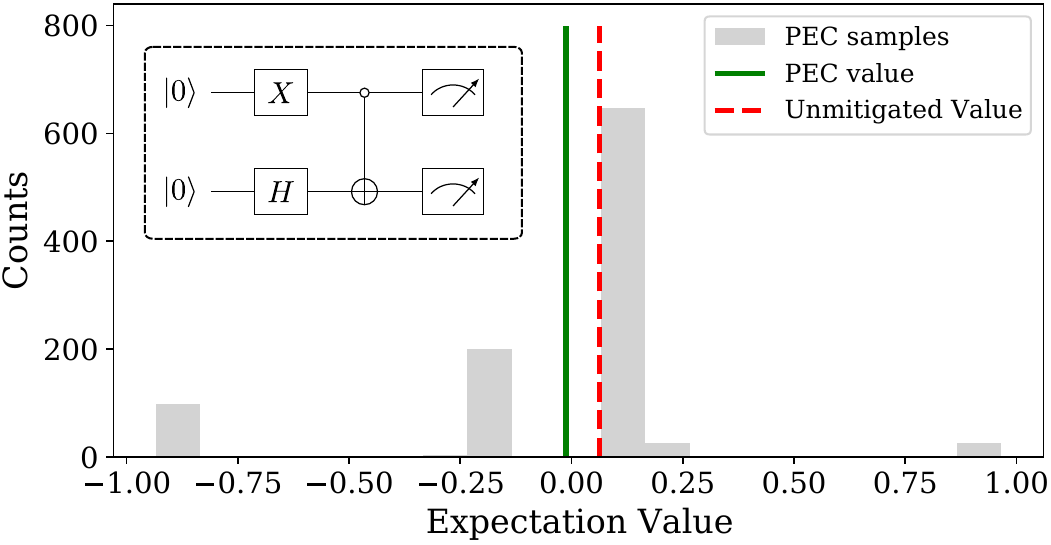}
    \caption{Expectation value estimated with PEC (green line) and the corresponding histogram of raw unbiased  PEC samples (gray bars). For comparison, also the unmitigated expectation value is shown (red dashed line). The exact ideal result is zero. For the numerics, we used a density matrix simulation and therefore shot noise is absent. The total number of PEC samples, corresponding to the sum of all the histogram counts, is 1000. The mean of all the unbiased PEC samples (i.e.\ the mean value of the gray histogram) corresponds to the error-mitigated PEC value represented as a vertical green line.  The circuit used in this example is shown in the inset.}\label{fig:pec_hist}
\end{figure}

We finally consider a toy example where Mitiq is used to apply probabilistic error cancellation. Consider the simple two-qubit circuit shown in the inset of \cref{fig:pec_hist}, corresponding to  $\mathcal U = \text{CNOT}_{1,2} \circ {X}_1 \circ  {H}_2$ (where $H_2$ is the Hadamard gate applied on the second qubit)\footnote{In this notation, the chronological order of the gates is from right to left, i.e., $\text{CNOT}$ is the last gate of $\mathcal U$.}. Assume that we want to measure the expectation value of $\mathcal O = |00\rangle \langle 00|$, whose exact theoretical value is zero. We also assume that each gate of the (simulated) backend is followed by local (single-qubit) depolarizing noise with error probability $p=0.1$.
Because of such noise, the unmitigated expectation value is nonzero ($0.0622$). However, after using Mitiq to implement PEC, one can improve the estimate by almost an order of magnitude ($0.0071$). The results are reported in \cref{fig:pec_hist}, where the histogram of the raw PEC samples is also visible.

\section{Zero-noise extrapolation module}\label{sec:zne}

We now describe the Mitiq library in more detail. The module structure is shown in \cref{fig:mitiq-tree} and includes a module to interface with supported quantum programming frameworks, several modules associated to different error mitigation techniques, and a module for benchmarking such techniques.

In this section, we focus on the zero-noise extrapolation module \inlinecode{mitiq.zne}, while  other error mitigation modules are considered in the next sections.

Zero-noise extrapolation was first introduced in~\cite{Temme_2017_PRL,Li_2017_PRX} and works by intentionally increasing (scaling) the noise of a quantum computation to then extrapolate back to the zero-noise limit. More specifically, let $\rho$ be a state prepared by a quantum circuit and $E^\dagger = E$ be an observable. We wish to estimate $\text{Tr} [\rho E] \equiv \langle E \rangle$ as though we had an ideal (noiseless) quantum computer, but there is a base noise level $\gamma_0$ which prevents us from doing so. For example, $\gamma_0$ could be the strength of a depolarizing channel in the circuit. The idea of zero-noise extrapolation is to compute
\begin{equation}\label{eqn:zne-scaling}
    \langle E(\gamma_i) \rangle =  \langle E (\lambda_i \gamma_0 ) \rangle
\end{equation}
where (real) coefficients $\lambda_{i} \ge 1$ scale the base noise $\gamma_0$ of the quantum computer. After this, a curve is fit to the data collected via Eq.~\eqref{eqn:zne-scaling} which is then extrapolated to the zero-noise limit. This produces an estimate of the noiseless expectation value $\langle E \rangle$.

To implement zero-noise extrapolation, we thus need two subroutines:
\begin{enumerate}
    \item A means of scaling the noise $\gamma_i = \lambda_i \gamma_0$ for different scale factors $\lambda_i$, and
    \item A means of fitting a curve to the noisy expectation values and extrapolating to the zero-noise limit.
\end{enumerate}
In the remainder of this section, we describe how these subroutines are implemented in Mitiq, showing several methods for both noise scaling as well as fitting/extrapolation, which we also refer to as inference.

\subsection{Noise scaling}\label{sec:noise-scaling}

In one of the first formulations of zero-noise extrapolation~\cite{Temme_2017_PRL}, noise is scaled in superconducting processors by implementing pulses at lower amplitudes for longer time intervals. Considering that most quantum programming languages support gate-model circuits and not pulse-level access, it can be convenient to scale noise in a manner which acts on unitary gates instead of underlying pulses. For this reason, Mitiq implements \textit{unitary folding}, introduced in~\cite{Giurgica_2020_arxiv}, as a noise scaling method.

\subsubsection{Unitary Folding}\label{sec:unitary-folding}

Unitary folding works by mapping gates (or groups of gates) $G$ to
\begin{equation} \label{eqn:unitary-folding}
    G \mapsto G G^\dagger G .
\end{equation}
This leaves the ideal effect of the circuit invariant but increases its depth. If $G$ is a gate of the circuit, we refer to the process as \textit{local folding}. If $G$ is the entire circuit, we call it \textit{global folding}.

In Mitiq, folding functions input a circuit and a \textit{scale factor} --- i.e., a number to increase the depth of the circuit by. (In Eq.~\eqref{eqn:zne-scaling}, each coefficient $\lambda_i$ is a scale factor.) The minimum scale factor is one (which corresponds to folding no gates), a scale factor of three corresponds to folding all gates, and scale factors beyond three fold some or all gates more than once.

For local folding, there is a degree of freedom for which gates to fold first. This order in which gates are folded can affect how the noise is scaled and thus the overall effectiveness of zero-noise extrapolation. Because of this, Mitiq defines several local folding functions in \inlinecode{mitiq.zne.scaling}, including:

\begin{enumerate}
    \item \inlinecode{fold\_gates\_from\_left}
    \item \inlinecode{fold\_gates\_from\_right}
    \item \inlinecode{fold\_gates\_at\_random}
\end{enumerate}

We explain how these functions work with the following example. We first define a circuit, here in Cirq, which for simplicity creates a Bell state.

\begin{lstlisting}[language=Python, caption={Defining a Bell state circuit in Cirq to be folded.}, firstnumber=1, label=lst:bell-state-circuit]
import cirq

qreg = cirq.LineQubit.range(2)
circ = cirq.Circuit(
    cirq.ops.H.on(qreg[0]), 
    cirq.ops.CNOT.on(qreg[0], qreg[1]),
)
print("Original circuit:", circ, sep="\n")
# Original circuit:
# 0: ---H---@---
#           |
# 1: -------X---
\end{lstlisting}

We can now use a local folding function, e.g. \inlinecode{fold\_gates\_from\_left}, to fold this circuit.
\begin{lstlisting}[language=Python, caption={Local folding from left on a Cirq circuit.}, firstnumber=13, label=lst:fold-left]
from mitiq.zne import scaling

folded = scaling.fold_gates_from_left(
    circ, scale_factor=2,
)
print("Folded circuit:", folded, sep="\n")
# Folded circuit:
# 0: ---H---H---H---@---
#                   |
# 1: ---------------X---
\end{lstlisting}
We see that the first Hadamard gate $H$ has been transformed as $H \mapsto H H^\dagger H$, to scale the depth of the circuit by a factor of two.

In Mitiq, folding functions do not modify the input circuit. Because of this, we can input the same circuit to \inlinecode{fold\_gates\_from\_right} to see the effect of this method.
\begin{lstlisting}[language=Python, caption={Local folding from right on a Cirq circuit. The \inlinecode{scaling} module is imported in \cref{lst:fold-left}.}, firstnumber=23]
folded = scaling.fold_gates_from_right(
    circ, scale_factor=2,
)
print("Folded circuit:", folded, sep="\n")
# Folded circuit:
# 0: ---H---@---@---@---
#           |   |   |
# 1: -------X---X---X---
\end{lstlisting}
Here, we see that the second (CNOT) gate is folded instead of the first (Hadamard) gate, as expected when we start folding from the right (or end) of the circuit instead of the left (or start) of the circuit.

The previous functions fold gates according to the following rules:

\begin{enumerate}
    \item If the scale factor is an odd integer
          $1 + 2n$, all gates are folded $n$ times.

    \item A generic real scale factor can always be written as $\lambda =1 + 2(n + \delta) $, where $n$ is an integer and $\delta<1$. In this case, all gates are folded $n$ times and, moreover, a subset of gates is folded one more time to better approximate the scale factor. The choice of this subset of gates can be random (in \inlinecode{fold\_gates\_at\_random}) or deterministic (in \inlinecode{fold\_gates\_from\_left} and \inlinecode{fold\_gates\_from\_right}).
\end{enumerate}

We emphasize that, although these examples used a Cirq \inlinecode{Circuit}, circuits can be defined in any supported quantum programming language and the interface is the same as above. In addition to Cirq, Mitiq supports other quantum libraries as listed in \cref{tab:backends}. By default, all folding functions return a circuit with the same type as the input circuit.

In the previous examples, each folded gate counts equally in the folded circuit depth. However, this may not be a reasonable assumption for realistic hardware as different gates have different noise levels. Because of this, each folding function in Mitiq supports ``folding by fidelity.'' This works by passing an input dictionary of gate fidelities (either known or estimated) as an optional argument to a folding function. More details on folding by fidelity can be found in \href{https://mitiq.readthedocs.io/en/latest/guide/guide_04-folding.html#folding-gates-by-fidelity}{Mitiq's documentation}.

Finally, we mention global folding. In contrast to local folding which folds subsets of gates, global folding folds the entire circuit until the input scale factor is reached. Below we show an example of global folding using the same Bell state circuit \inlinecode{circ} defined in \cref{lst:bell-state-circuit}.

\begin{lstlisting}[language=Python, caption={Global folding on a Bell state circuit.}, firstnumber=1]
folded = scaling.fold_global(circ, scale_factor=3.0)
print("Folded circuit:", folded, sep="\n")
# Folded circuit:
# 0: ---H---@---@---H---H---@---
#           |   |           |
# 1: -------X---X-----------X---
\end{lstlisting}

\noindent Here, we see that the entire Bell state circuit has been folded once to reach the input scale factor of three. If the input scale factor is not reached by an integer number of global folds, \inlinecode{fold\_global} will fold a group of gates from the end of the circuit such that the scale factor is reached.

\subsubsection{Parameter-noise scaling}\label{sec:parameter-noise}

A gate is an abstract elementary operation which, however, is physically implemented as a continuous dynamical evolution. This evolution is generated by a suitable time-dependent control of a Hamiltonian that depends on the details of the hardware. Errors in the calibration of control pulses (e.g.~pulse-area errors) or the classical noise affecting their implementation (e.g.~electronic noise) can generate a dynamical channel which is different from the desired ideal gate.

In order to mitigate these type of errors, we need a practical way of scaling them.
In principle this would require the detailed knowledge of the platform-dependent pulses and Hamiltonians, however, in Mitiq a simplified noise model is used instead. The simplified model is based on the fact that any unitary gate $G$ can always be expressed as $G=\exp(-i H)$, for some constant Hamiltonian $H=H^\dag$ (which may be different from the physical one).
Therefore, each  unitary gate admits a natural parametrization with respect to a real exponent $\theta$:
\begin{equation} \label{eq:g_of_theta}
    G(\theta) = \exp(-i H \theta) = G^\theta.
\end{equation}
A multi-parameter version of Eq.~\eqref{eq:g_of_theta} was considered in~\cite{Giurgica_2020_arxiv}, but is currently not used in Mitiq.
It is also worth to mention that gates
are often directly defined in the parametric form of Eq.~\eqref{eq:g_of_theta} as, for example, in the case of Pauli rotations.

In this setting, a noise model approximately modeling calibration and control errors can be expressed with respect to the classical parameter $\theta$. We can assume that the actual gate is generated by a noisy parameter $\hat \theta$ that we can model as a random variable with mean $\theta$ and with some variance $\sigma^2$. Noise scaling can be achieved by artificially injecting additional classical noise:
\begin{equation} \label{eq:classical_noise}
    \hat \theta \rightarrow \hat \theta^{(\lambda)} =  \hat \theta + \hat \delta
\end{equation}
where $\hat \delta$ is a random variable with zero mean and variance equal to $(1 - \lambda) \sigma^2$, such that the resulting noise scaled parameter $\hat \theta^{(\lambda)}$ has mean $\theta$ and variance $\lambda \sigma^2$.

In practice, if $\sigma^2$ is known for each noisy gate, parameter scaling can be obtained by randomly over-rotating or under-rotating each gate according to the stochastic angles defined in Eq.~\eqref{eq:g_of_theta}.
This noise scaling technique can be applied with Mitiq as shown in the next Codeblock.

\begin{lstlisting}[language=Python, caption={Applying parameter-noise
scaling to a quantum circuit. The same base level of noise  (\inlinecode{base\_variance}) is assumed
for each gate of the circuit.}, firstnumber=1, label=cb_parameter]
from mitiq.zne.scaling import scale_parameters

scaled_circuit = scale_parameters(
    circuit=<the circuit to scale>,
    scale_factor=<the noise scale factor>,
    base_variance=<the base level of noise>,
)

\end{lstlisting}

If the value of the base noise $\sigma^2$ is unknown, it needs to be estimated in order to apply this noise scaling method.
The function  \inlinecode{compute\_parameter\_variance} in the sub-module \inlinecode{mitiq.zne.scaling} can be used for this task.
Alternatively, the user may independently perform a custom estimation of $\sigma^2$ and only use Mitiq for the noise scaling step described in \cref{cb_parameter}.

The full application of ZNE obtained via the parameter-noise scaling method is shown in the next Codeblock.

\begin{lstlisting}[language=Python, caption={Applying parameter-noise
scaling for ZNE.}, firstnumber=1, label=parameter_zne]
from functools import partial
from mitiq.zne.scaling import compute_parameter_variance, scale_parameters

base_variance = compute_parameter_variance(
    executor, <gate>, <qubit>,
)
scale_param_noise = partial(
    scale_parameters, base_variance=base_variance,
)
zne_value = zne.execute_with_zne(
    circuit,
    executor,
    scale_noise=scale_param_noise,
    num_to_average=10,
)
\end{lstlisting}
\subsubsection{Using noise scaling methods in \inlinecode{execute\_with\_zne}}

As mentioned in \cref{sec:main-usage}, the default noise scaling method in \inlinecode{execute\_with\_zne} is \inlinecode{fold\_gates\_at\_random}. Different methods can be used by passing an optional argument to \inlinecode{execute\_with\_zne}. For example, to use global folding, one can do the following.
\begin{lstlisting}[language=Python, caption={Using zero-noise extrapolation with global folding by passing \inlinecode{fold\_global} as an optional argument to \inlinecode{execute\_with\_zne}. The \inlinecode{circuit} and \inlinecode{executor} are as in \cref{sec:main-usage}.}, firstnumber=1]
from mitiq.zne import execute_with_zne
from mitiq.zne.scaling import fold_global

zne_value = execute_with_zne(
    circuit,
    executor,
    scale_noise=fold_global,
)
\end{lstlisting}
Depending on the noise model of the quantum processor, using a different folding method may better scale the noise and lead to better results.

To end the discussion on noise scaling, we note that some scaling methods are deterministic while some are non-deterministic. In particular, global folding and local folding from left/right return the same folded circuit if the scale factor is the same, but \inlinecode{fold\_gates\_at\_random} can return different circuits for the same scale factor. Because of this, the function \inlinecode{execute\_with\_zne} has another optional argument \inlinecode{num\_to\_average} which corresponds to the number of times to compute expectation values at the same scale factor. For example, if \inlinecode{num\_to\_average = 3}, the noise scaling method is called three times at each scale factor, and the expectation value at this scale factor is the average over the three runs. Such averaging can smooth out effects due to non-deterministic noise scaling and lead to better results in zero-noise extrapolation. \cref{fig:h2-energy-surface}(b) uses \inlinecode{fold\_gates\_at\_random} with \inlinecode{num\_to\_average = 5}.

\subsection{Classical inference: Factory objects}\label{sec:factories}

In Mitiq, a \inlinecode{Factory} object is a self-contained representation of a classical inference technique. In effect, it performs the ``extrapolation'' part of zero-noise extrapolation.
This representation is hardware-agnostic and even quantum-agnostic since it only deals with classical data --- namely, the input and output of a noisy computation. The main tasks of a factory are as follows:

\begin{enumerate}
    \item Compute the expectation value by running an executor function at a given noise level, and record the result;

    \item Determine the next noise level at which the expectation value should be computed;

    \item Perform classical inference using the history of noise levels and expectation values to compute the zero-noise extrapolated value.
\end{enumerate}

The structure of a \inlinecode{Factory} is designed to account for adaptive fitting techniques in which the next noise level depends on the history of previous noise levels and expectation values. In Mitiq, (adaptive) fitting techniques in zero-noise extrapolation are represented by specific factory objects. All built-in factories, summarized in \cref{tab:factories}, can be imported from the \texttt{mitiq.zne.inference} module.

\begin{table}[ht]
    \centering\begin{tabular}{r p{9cm}}
        \textbf{Class}             & \textbf{Extrapolation Method}                                                     \\
        \toprule
        \texttt{LinearFactory}     & Extrapolation with a linear fit.                                                  \\
        \texttt{RichardsonFactory} & Richardson extrapolation.                                                         \\
        \texttt{PolyFactory}       & Extrapolation with a polynomial fit.                                              \\
        \texttt{ExpFactory}        & Extrapolation with an exponential fit.                                            \\
        \texttt{PolyExpFactory}    & Similar to \texttt{ExpFactory} but the exponent can be a non-linear polynomial.   \\
        \texttt{AdaExpFactory}     & Similar to \texttt{ExpFactory} but the noise scale factors are adaptively chosen. \\
    \end{tabular}
    \caption{Factories that can be imported from \inlinecode{mitiq.zne.inference} to perform different extrapolation methods. More information is available in the \href{https://mitiq.readthedocs.io/en/latest/guide/guide_05-factories.html}{Mitiq documentation} and an analysis of the different extrapolation methods can be found in Ref.~\cite{Giurgica_2020_arxiv}.}\label{tab:factories}
\end{table}

\subsubsection{Using factories in \inlinecode{execute\_with\_zne} to perform different extrapolation methods}

We now show examples of performing zero-noise extrapolation with fitting techniques defined by factories in \cref{tab:factories}. As mentioned in \cref{sec:main-usage}, this is done by providing a factory as an optional argument to \inlinecode{execute\_with\_zne}. To instantiate a non-adaptive factory, we input the noise scale factors we want to compute the expectation values at, as shown below for the \inlinecode{LinearFactory}.
\begin{lstlisting}[language=Python, caption={Initializing a factory object.}, firstnumber=1]
from mitiq.zne.inference import LinearFactory

linear_factory = LinearFactory(scale_factors=[1.0, 2.0, 3.0])
\end{lstlisting}
Here the \inlinecode{scale\_factors} define the noise levels at which to compute expectation values during zero-noise extrapolation. This factory can now be used as an argument in \inlinecode{execute\_with\_zne} as follows. As in \cref{sec:main-usage}, the \inlinecode{circuit} is the quantum program which prepares a state of interest and the \inlinecode{executor} is a function which executes the circuit and returns the expectation value of an observable.
\begin{lstlisting}[language=Python, caption={Using a factory object as an optional argument of \inlinecode{mitiq.zne.execute\_with\_zne}.}, firstnumber=6]
from mitiq.zne import execute_with_zne

zne_value = execute_with_zne(
    circuit,
    executor,
    factory=linear_factory,
)
\end{lstlisting}
Instead of the default Richardson extrapolation at noise scale factors $1, 2$ and $3$, this call to \inlinecode{execute\_with\_zne} will perform linear extrapolation at the specified noise scale factors. As mentioned in \cref{sec:noise-scaling}, different noise scaling methods can also be used with the optional argument \inlinecode{scale\_noise}.

Most extrapolation techniques implemented in Mitiq are static (i.e., non-adaptive) and can be instantiated in a similar manner as the \inlinecode{LinearFactory}. For example, to use a second-order polynomial fit, we use a \inlinecode{PolyFactory} object as follows.
\begin{lstlisting}[language=Python, caption={Instantiating a second-order \inlinecode{PolyFactory}.}, firstnumber=1]
from mitiq.zne import execute_with_zne
from mitiq.zne.inference import PolyFactory

zne_value = execute_with_zne(
    circuit,
    executor,
    factory=PolyFactory(scale_factors=[1.0, 2.0, 3.0], order=2),
)
\end{lstlisting}
Other static factories follow similar patterns but may have additional arguments in their constructors. For example, \inlinecode{ExpFactory} can take in a value for the horizontal asymptote of the exponential fit. For full details, see the \href{https://mitiq.readthedocs.io/en/latest/guide/guide_05-factories.html}{Mitiq documentation}.

Last, we show an example of an adaptive fitting technique defined by the \inlinecode{AdaExpFactory}. To use this method (introduced and described in Ref.~\cite{Giurgica_2020_arxiv}), we can do the following:
\begin{lstlisting}[language=Python, caption={Using \inlinecode{execute\_with\_zne} with an adaptive fitting technique.}, firstnumber=1]
from mitiq.zne import execute_with_zne
from mitiq.zne.inference import AdaExpFactory

zne_value = execute_with_zne(
    circuit,
    executor,
    factory=AdaExpFactory(scale_factor=2.0, steps=5),
)
\end{lstlisting}
Instead of a list of scale factors, here we provide the initial scale factor and the rest are determined adaptively. The number of scale factors determined is equal to the argument \inlinecode{steps}. Additional arguments which can be passed into the \inlinecode{AdaExpFactory} are described in the \href{https://mitiq.readthedocs.io/en/latest/apidoc.html#mitiq.factories.AdaExpFactory}{Mitiq documentation}.

\subsubsection{Using custom fitting techniques}

A custom fitting technique can be used in Mitiq by defining a new factory class which inherits from the abstract class \inlinecode{mitiq.zne.inference.Factory} (for general techniques) or \inlinecode{mitiq.zne.inference.BatchedFactory} (a subclass of \inlinecode{Factory} suitable for non-adaptive techniques). To get noise scale factors and expectation values, the methods \\
\inlinecode{Factory.get\_scale\_factors()} and \inlinecode{Factory.get\_expectation\_values()} can be used.

Below, we define a static factory which performs a second-order polynomial fit and forces the expectation value to be in the interval $[-1, 1]$.
\begin{lstlisting}[language=Python, caption={Defining a custom fitting technique by creating a new factory object.}, firstnumber=1]
from mitiq.zne.inference import (
    BatchedFactory, PolyFactory,
)
import numpy as np

class MyFactory(BatchedFactory):
    @staticmethod
    def extrapolate(
        scale_factors, exp_values, full_output,
    ):
        result = PolyFactory.extrapolate(
            scale_factors,
            exp_values,
            order=2,
            full_output=full_output,
    )
        if not full_output:
            return np.clip(result, -1, 1)
        if full_output:
            # In this case "result" is a tuple
            zne_limit = np.clip(results[0], -1, 1)
            return (zne_limit, *results[1:])
\end{lstlisting}
This factory can now be used as an argument in \inlinecode{execute\_with\_zne} to use the custom fitting technique.
Other fitting techniques can be defined in a similar manner as the code block above.

\section{Probabilistic error cancellation module}\label{sec:pec}

\begin{figure}
    \centering
    \includegraphics[width=0.75\textwidth]{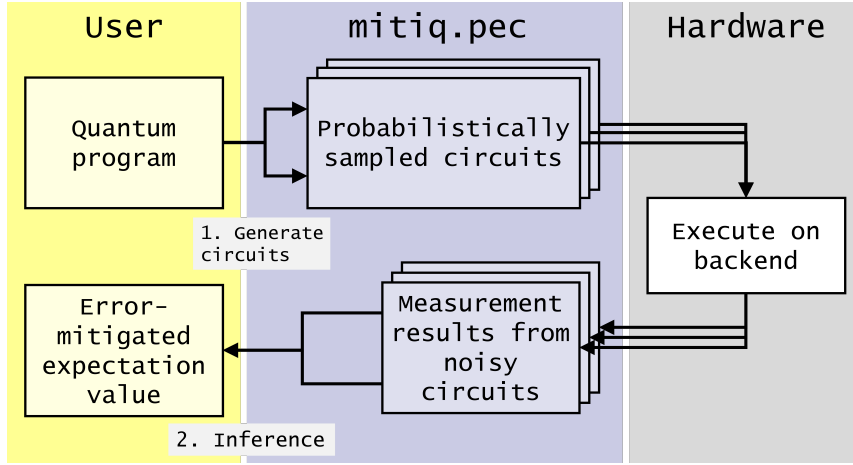}
    \caption{Overview of the probabilistic error cancellation workflow in Mitiq. An input quantum program is converted into a set of auxiliary circuits which are probabilistically sampled according to the PEC technique described in the main text. These sampled circuits are executed on the back-end according to a user-defined executor function and produce a set of noisy results. The noisy expectation values are combined (with a  suitable linear combination) to obtain an unbiased estimate of the ideal expectation value.}\label{fig:pec-pipeline}
\end{figure}

Probabilistic error cancellation (PEC)~\cite{Temme_2017_PRL, endo2018practical} is another error mitigation technique which is available in Mitiq. Its workflow is schematically represented in \cref{fig:pec-pipeline}: a set of auxiliary circuits are probabilistically sampled, executed on a noisy backend and, eventually, the noisy results are post-processed to infer an error-mitigated expectation value.
In principle, this method can probabilistically remove the noise of a quantum computer without additional resources apart from a higher sampling overhead. More information about the advantages and the limitations of PEC is given in \cref{subsec:pec-limitations}.

A key step of PEC is to represent each ideal unitary gate $\mathcal G$ in a circuit as an average over a set of noisy gates which are physically
implementable $\{ \mathcal O_\alpha \}$, weighted by a real quasi-probability distribution $\eta(\alpha)$:

\begin{equation}\label{eq:quasi-dist}
    \mathcal{G} = \sum_\alpha \eta(\alpha) \mathcal O_\alpha,
\end{equation}
where $\sum_\alpha \eta(\alpha) = 1$ (trace-preserving condition). The calligraphic operators $\mathcal G$ and $\{\mathcal O_\alpha\}$ should be considered as linear super-operators acting on density matrices and not on state vectors~\cite{Temme_2017_PRL, endo2018practical}.
If a representation like Eq.~\eqref{eq:quasi-dist} is known for each ideal gate of a circuit, then any ideal expectation value can be
estimated as a Monte Carlo average over different noisy circuits, each one sampled according to the quasi-probability distributions associated to
the ideal gates~\cite{Temme_2017_PRL, endo2018practical}. The real coefficients $\eta(\alpha)$ which appear in Eq.~\eqref{eq:quasi-dist}
can be negative for some values of $\alpha$ and, because of this negativity, the required number of Monte Carlo samples can be
large~\cite{Temme_2017_PRL, endo2018practical}. In principle, assuming a perfect tomographic knowledge of the noisy gates $\mathcal O_\alpha$,
this method allows for a perfect cancellation of the hardware noise (for a sufficiently large number of samples).

In the remainder of this section, we describe how one can define gate representations and how one can probabilistically sample from them using Mitiq.

\subsection{Noisy Operations}

The r.h.s.~of Eq.~\eqref{eq:quasi-dist} is a sum over noisy operations $\mathcal O_\alpha$.
A noisy operation is an elementary gate (or a small sequence of gates) acting on specific qubits
which can be physically implemented on hardware. To each noisy operation we can associate a (small)
\inlinecode{QPROGRAM} describing the gates to be applied on the physical qubits. Moreover, from a
quantum tomography analysis, one can associate to a noisy operation also a numerical matrix representing the completely-positive
and trace-preserving channel induced by the operation.
In Mitiq, this concept is captured by the \inlinecode{NoisyOperation} class, which can be
initialized as follows:

\begin{lstlisting}[language=Python, 
caption={Initialization of a \inlinecode{NoisyOperation} from a
sequence of operations and the (optional) super-operator matrix. }, 
firstnumber=1]
from mitiq.pec.types import NoisyOperation

noisy_operation = NoisyOperation(
    circuit=<sequence of operations as a QPROGRAM>, 
    channel_matrix=<optional superoperator matrix>,
)
\end{lstlisting}

Once the set of all noisy operations $\{\mathcal O_\alpha \}$ has been defined, we can associate to each operation the
corresponding quasi-probability $\eta(\alpha)$ via a simple Python dictionary:

\begin{lstlisting}[language=Python, 
caption={Defining a basis expansion as a Python dictionary which associates  a real coefficient to each noisy operation.}, firstnumber=7, label=lst:basis_expansion]
basis_expansion = {
    <1st noisy operation>: <1st real coefficient>,
    <2nd noisy operation>: <2nd real coefficient>,
    ...
}
\end{lstlisting}

\subsection{\texttt{OperationRepresentation} Objects}\label{sec:representations}

The dictionary in the previous \cref{lst:basis_expansion} completely defines the linear combination in the r.h.s.~of Eq.~\eqref{eq:quasi-dist} but it
contains no information about the l.h.s.~of Eq.~\eqref{eq:quasi-dist}. This motivates the use of an \inlinecode{OperationRepresentation}
class which can be used to store and manipulate all the information which is contained in Eq.~\eqref{eq:quasi-dist}.

\begin{lstlisting}[language=Python, 
caption={Initializing an \inlinecode{OperationRepresentation} object. The first argument is the  ideal operation that we want to express as a linear combination of noisy operations. The second argument is the associated \inlinecode{basis\_expansion} which can be defined as shown in \cref{lst:basis_expansion}.}, firstnumber=12, label=lst:operation-representation]
from mitiq.pec.types import OperationRepresentation

operation_rep = OperationRepresentation(
    ideal=<ideal operation as a QPROGRAM>,
    basis_expansion=<basis expansion dictionary>,
)
\end{lstlisting}

Given a list of \inlinecode{OperationRepresentation} objects, associated to all the gates of a circuit of interest, the user can easily apply PEC via the function \inlinecode{execute\_with\_pec} as shown in \cref{lst:execute_with_pec} of \cref{sec:getting-started}.

\subsection{How to determine the quasi-probability representations?}

In practice, depending on how detailed is the knowledge of the hardware noise model, there are two main ways of deriving quasi-probability representations for PEC.

\noindent \emph{Method 1:} If the hardware noise model is well approximated by a simplified theoretical quantum channel (e.g.\ depolarizing or amplitude damping), one can typically apply known analytical expressions to compute the quasi-probability representations of arbitrary gates~\cite{Temme_2017_PRL}.

\noindent \emph{Method 2:} Assuming an over-simplified noise model may be a bad approximation. In this case, the suggested approach is to perform the complete process tomography of a basis set of implementable noisy operations (e.g.\ the native gate set of the backend). Given the superoperators of the noisy implementable operations, one can obtain the quasi-probability representations as solutions of numerical optimization problems~\cite{Temme_2017_PRL}. In Mitiq, this is possible through the \inlinecode{find\_optimal\_representation()} function that can
be imported from \inlinecode{mitiq.pec.representations}. An example showing how to use this function is given in the section called \textit{What additional options are available in PEC?} of the Mitiq documentation (see \cref{sec:documentation}).

\subsection{Sampling Functions}
The function \inlinecode{execute\_with\_pec} internally performs the Monte Carlo sampling process
which is necessary to estimate an expectation value with PEC.
However, the user may be interested in manually sampling gates and circuits for a variety of reasons,
{\it e.g.}, for research purposes, for intermediate manipulations, for efficiency optimizations, etc.

In particular, to sample an implementable \inlinecode{NoisyOperation} from the quasi-probability distribution of an ideal operation one can do as follows:

\begin{lstlisting}[language=Python, 
caption={Sampling an implementable \inlinecode{NoisyOperation} from the quasi-probability representation of an ideal operation. The quasi-probability representation is given by the  \inlinecode{OperationRepresentation} object defined in \cref{lst:operation-representation}. In addition to the sampled \inlinecode{noisy\_operation}, the method \inlinecode{sample()} returns the associated coefficient (\inlinecode{eta}) that appears in Eq.~\eqref{eq:quasi-dist} and its sign (\inlinecode{sign}).},
firstnumber=18, label=lst:sample_sequence]
noisy_operation, sign, eta = operation_rep.sample()
\end{lstlisting}

Typically, one is interested in sampling an entire implementable circuit from the quasi-probability representation of an ideal circuit. This can be easily achieved via the\\\inlinecode{sample\_circuit} function, which internally performs repeated calls to the previous\\\inlinecode{sample\_sequence} function:

\begin{lstlisting}[language=Python, 
caption={Sampling an implementable circuit from the quasi-probability representation of an \inlinecode{ideal\_circuit}. Such quasi-probability distribution is implicitly deduced from the input list of \inlinecode{OperationRepresentations} objects associated to the gates of the input  \inlinecode{ideal\_circuit} .},
firstnumber=19, label=lst:sample_circuit]
from mitiq.pec.sampling import sample_circuit

sampled_circuit, sign, norm = sample_circuit(
	ideal_circuit=<ideal circuit as a QPROGRAM>,
	representations=<list of oper. representations>,
)
\end{lstlisting}

\section{Clifford data regression module}\label{sec:cdr}
In this section, we present the \inlinecode{mitiq.cdr} module
which implements two recent error mitigation approaches known as Clifford data regression (CDR) and variable noise Clifford data regression (vnCDR)~\cite{czarnik2020error, lowe2020unified}. In both techniques, a trained regression model mapping noisy to exact expectation values is used to mitigate the effect of noise on some observable of interest. The model is trained using data produced by the execution of near-Clifford circuits performed on a noisy quantum computer and on a classical simulator.

\begin{figure}[t]
    \centering
    \includegraphics[width=0.35\columnwidth]{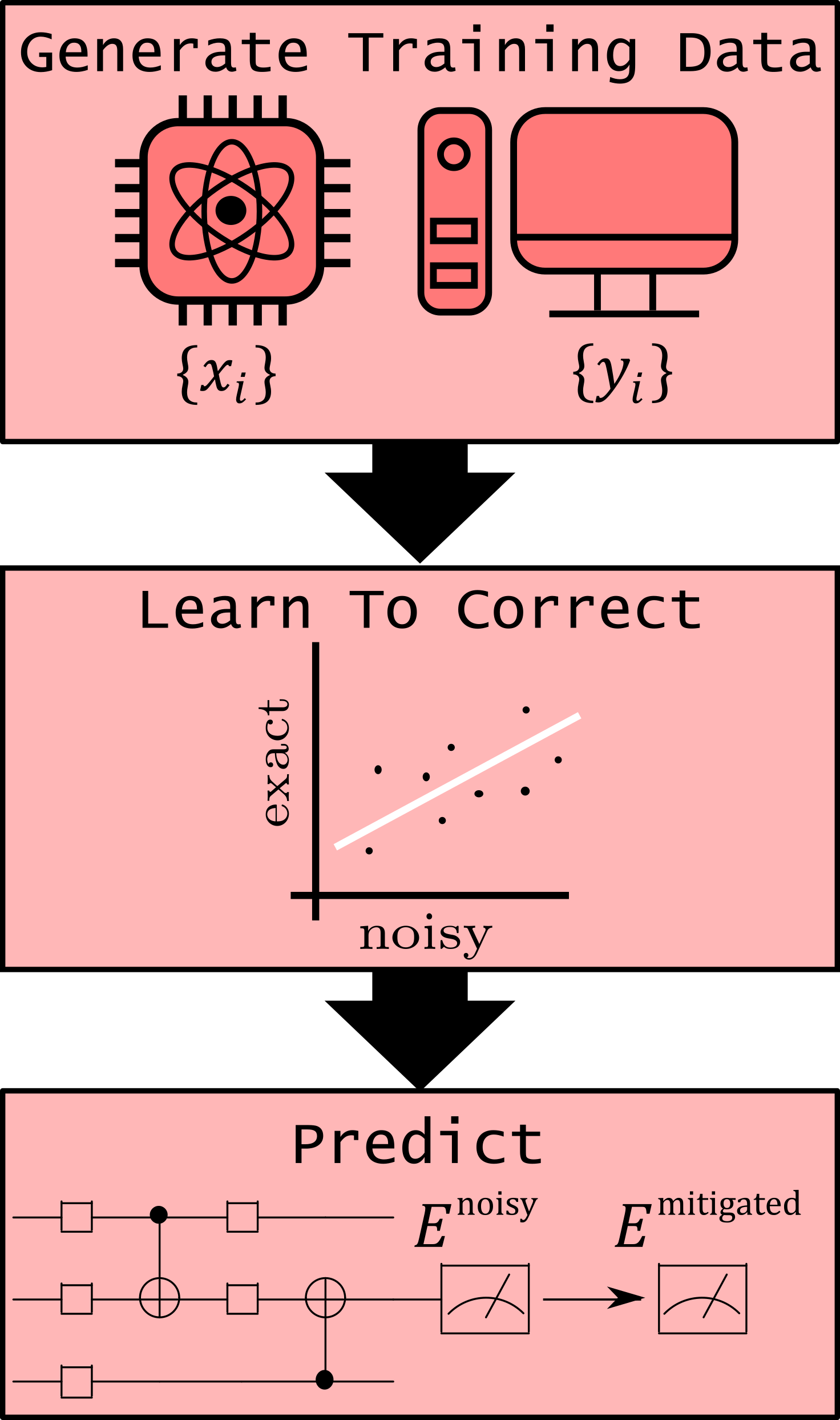}
    \caption{Summary of the CDR mitigation method showing the steps realized to mitigate an observable of interest. First the training set composed of noisy $\{x_{i}\}$ and exact $\{y_{i}\}$ expectation values is generated using near-Clifford circuits which are classically simulable. This data is used to fit a linear ansatz which is then used to estimate the noise-free value for some observable of interest $E$. We can visualize vnCDR as adding another axis to the training data, along which noise is increased. Diagram modified from~\cite{czarnik2020error}.}\label{fig:cdr_diagram}
\end{figure}

\subsection{Clifford data regression (CDR)}

The Clifford data regression~\cite{czarnik2020error} technique uses near-Clifford quantum circuit data to learn a model approximating effects of the noise on an expectation value of an observable $\langle E\rangle = \textrm{Tr} \rho E$ for a quantum state $\rho$ given by a quantum circuit of interest. The learned model is used  to mitigate the noisy expectation value $\langle E(\gamma_0)\rangle$ obtained with  a  quantum computer with the base noise level $\gamma_0$. The mitigated expectation value $\langle E \rangle^{\rm mitigated}$ is obtained using the following procedure:
\begin{enumerate}
    \item Construct the training circuits corresponding to states $\{ \rho^{\rm train}_i, i=1,\dots,n \}$ by replacing non-Clifford gates in the circuit of interest by Clifford gates.
    \item For each training circuit $\rho^{\rm train}_i$ evaluate classically a noiseless expectation value of $E$,  $y_i =  \textrm{Tr} \rho_i E$,  and its noisy expectation value $x_i$ using a quantum computer.
    \item Fit exact and noisy expectation values of the  training circuits $\{(x_i,y_i)\}$  with a linear model $y = ax+b $.
    \item Use the fitted model to mitigate $\langle E(\gamma_0)\rangle$
          \begin{equation*}
              \langle E \rangle^{\text{mitigated}} = a \langle E(\gamma_0) \rangle + b.
          \end{equation*}
\end{enumerate}

\begin{figure*}
    \centering
    \includegraphics[width=.7\textwidth]{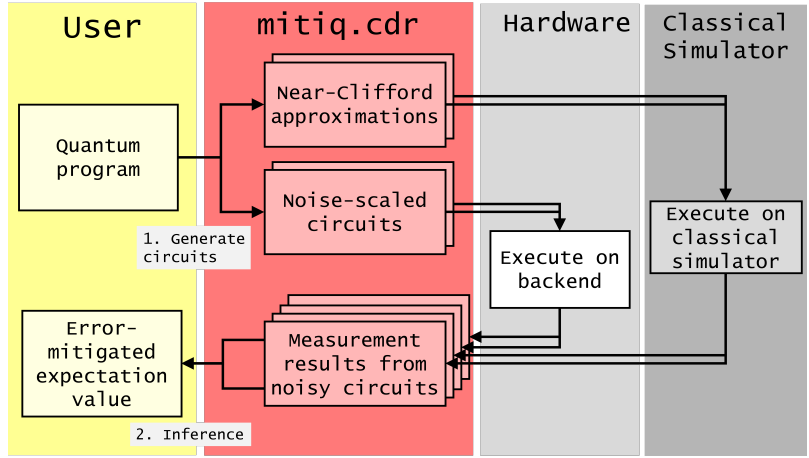}
    \caption{Overview of the Clifford data regression  workflow in Mitiq. An input quantum program is converted into a set of noise-scaled circuits and a set of training circuits (near-Clifford approximations in which the noise is also scaled). All the auxiliary circuits are executed on the real backend while the near-Clifford training circuits are executed also on a classical simulator. The noisy and the exact (simulated) results are post-processed to infer the ideal expectation value of the original quantum program.}\label{fig:cdr-pipeline}
\end{figure*}

\subsection{Variable noise Clifford data regression (vnCDR)}

CDR can be generalized to enable learning the noise effects from near-Clifford training circuits simulated at different noise levels $\lambda_l$.  This approach is called variable noise  Clifford data regression~\cite{lowe2020unified} and can be used to learn a zero-noise extrapolation model for an observable $E$ and a quantum circuit preparing the state $\rho$. The vnCDR procedure to obtain $\langle E \rangle^{\rm mitigated}$ includes evaluation of the training circuits on a quantum computer at different noise rates $\lambda_l \gamma_0$ and fitting a extrapolation model:
\begin{enumerate}
    \item Prepare the training circuits $\{ \rho^{\rm train}_i, i=1,\dots, n \}$ using Clifford substitutions, following the same procedure for CDR.
    \item For each training circuit $\rho^{\rm train}_i$ evaluate classically  a noiseless expectation value of $E$,  $y_i =  \textrm{Tr} \rho_i E$,  and its noisy expectation values $x_{i,l}$ using a quantum computer with several noise rates $\lambda_l \gamma_0$, $\lambda_l \ge 1, l = 1,\dots,m $.
    \item Fit the expectation values of the training circuits with a linear ansatz given by $y = f(x_1, x_2, \ldots, x_{m})$. Where
          \begin{equation}
              f(x_1, x_2, \dots, x_{m}) = \sum_{l=1}^{m} x_l a_l  + b\  .
          \end{equation}
    \item Use the fitted ansatz to correct the noisy expectation values  of $E$:
          \begin{displaymath}
              \langle E \rangle^{\rm mitigated} = f(\langle E(\lambda_1\gamma_0)\rangle, \langle E(\lambda_2\gamma_0)\rangle, \dots, \langle E(\lambda_{m}\gamma_0)\rangle)\ .
          \end{displaymath}
\end{enumerate}
The default linear ansatz used within Mitiq includes a constant term. Recently this was shown to lead to better mitigated results on real quantum hardware~\cite{sopena2021}.

\subsection{Applying CDR and vnCDR with Mitiq}

Clifford data regression is implemented in Mitiq according to the workflow schematically represented in \cref{fig:cdr-pipeline}.
This error mitigation technique can be applied with the following code Codeblock:

\begin{lstlisting}[language=Python, 
caption={Applying CDR with Mitiq. The function \inlinecode{execute\_with\_cdr} can be used to mitigate errors the expectation values of the input  \inlinecode{observables}. The input \inlinecode{executor} is a user-defined function for running the input circuit and the associated training circuits on a quantum backend. The input \inlinecode{simulator} is the ideal counterpart of the noisy \inlinecode{executor} and is
necessary to obtain exact classical simulations of the (near-Clifford) training circuits.}, firstnumber=1, label=cdr]
from mitiq.cdr import execute_with_cdr

cdr_values = execute_with_cdr(
    circuit,
    executor,
    simulator=<near-Clifford classical simulator>,
    observables=<the observables to mitigate>,
)
\end{lstlisting}

Similarly, variable-noise Clifford data regression can be applied by specifying the optional list of noise scale factors in the function \inlinecode{execute\_with\_cdr}.

\begin{lstlisting}[language=Python, 
caption={Applying vnCDR with Mitiq by calling \inlinecode{execute\_with\_cdr} and passing a list of noise \inlinecode{scale\_factors}. Optionally, a noise scaling method can be specified via the argument \inlinecode{scale\_noise}, whose default value is \inlinecode{fold\_gates\_at\_random}.}, firstnumber=1]
from mitiq.cdr import execute_with_cdr

cdr_values = execute_with_cdr(
    # The same arguments used for CDR:
    ...
    # Additional argument for applying vnCDR:
    scale_factors=<the noise scale factors>,
)
\end{lstlisting}

One of the key features of both CDR and vnCDR is the construction of a set of classically simulable near-Clifford circuits. At the time of this writing, CDR implemented within Mitiq assumes that the input circuit is pre-compiled in the following gate set $\{R_{Z}, \sqrt{X}, \text{CNOT}\}$. This ensures that all the non-Clifford gates are contained in the $R_{Z}$ gates. This is particularly suitable for IBM processors but may be less appropriate for other backends. Different gate sets may be supported in the future.

\section{Additional library information}\label{sec:software-details-library-information}
In this section, we provide technical details and meta-information about the Mitiq library.

\subsection{Alternative ways of using Mitiq}\label{sec:alternate-ways-to-use-zne}

As we have already shown, errors affecting the estimation of expectation values can be reduced with appropriate functions returning the mitigated expectation value as a real number, e.g.  \inlinecode{execute\_with\_zne}, \inlinecode{execute\_with\_pec}.
Here, we show two alternative methods for applying the same error mitigation process. Depending on context, these alternative but equivalent methods may provide a simpler usage.

The first method is provided by the function \inlinecode{mitigate\_executor} which inputs the same arguments as \inlinecode{execute\_with\_*} except the quantum circuit. This function returns a new executor which implements error mitigation when it is called with a quantum program, as shown below.
\begin{lstlisting}[language=Python, 
caption={Modifying an \inlinecode{executor} with the function \inlinecode{mitigate\_executor}. The new \inlinecode{mitigated\_executor} performs zero-noise extrapolation when called on a quantum circuit.}, firstnumber=1]
from mitiq.zne import mitigate_executor

mitigated_executor = mitigate_executor(
    executor,
    scale_noise=<noise scaling method>,
    factory=<inference method>,
)

zne_value = mitigated_executor(circuit)
\end{lstlisting}

The \inlinecode{mitigate\_executor} function can also be imported from other modules in order to apply different techniques. For example, probabilistic error cancellation can be applied after importing \inlinecode{mitigate\_executor} from  \inlinecode{mitiq.pec}.

The second method is to directly decorate the \inlinecode{executor} function such that it automatically performs error mitigation when called. Also in this case, one should use the decorator corresponding to the desired error mitigation technique, e.g.: \inlinecode{zne\_decorator},
\inlinecode{pec\_decorator}, etc.
\begin{lstlisting}[language=Python, 
caption={Decorating an \inlinecode{executor} with \inlinecode{zne\_decorator} so that zero-noise extrapolation is implemented when the \inlinecode{executor} is called on a quantum program}, firstnumber=1]
from mitiq import QPROGRAM
from mitiq.zne import zne_decorator

@zne_decorator(
    factory=<inference method>,
    scale_noise=<noise scaling method>,
)
def executor(circuit: QPROGRAM) -> float:
    ...

zne_value = executor(circuit)
\end{lstlisting}
In the above Codeblock, the \inlinecode{zne\_decorator} takes the same optional arguments as \\
\inlinecode{execute\_with\_zne}. If no optional arguments are used, the decorator should still be called with parentheses, e.g. \inlinecode{@zne\_decorator()}.

Decorators (or \inlinecode{mitigate\_executor}) could be used to easily  stack multiple error mitigation techniques. For example, in the next Codeblock, a noisy executor is first mitigated with PEC and later with ZNE.

\begin{lstlisting}[language=Python, 
caption={Multiple decorators can be used to combine different error mitigation methods}, firstnumber=1]
from mitiq import QPROGRAM
from mitiq.zne import zne_decorator
from mitiq.pec import pec_decorator

@zne_decorator(<zne arguments>)
@pec_decorator(<pec arguments>)
def executor(circuit: QPROGRAM) -> float:
    ...

mitigated_value = executor(circuit)
\end{lstlisting}
Whether there is any practical advantage in combining multiple techniques is still an open research question. Mitiq can be an appropriate toolkit for exploring this research direction.

\subsection{Mitiq documentation}\label{sec:documentation}

Mitiq's documentation is hosted online at \href{https://mitiq.readthedocs.io}{https://mitiq.readthedocs.io} and includes a User's Guide and an API glossary. The User's Guide contains more information on topics covered in this manuscript and additional information on topics not covered here; for example, more examples of executor functions and an advanced usage guide for factory objects. The API glossary is auto-generated from the docstrings (formatted comments to code objects) and contains information about public functions and classes defined in Mitiq.

\subsection{Contribution guidelines}\label{sec:development-practices}
We welcome contributions to Mitiq from the larger community of quantum software developers. Contributions can come in the form of feedback about the library, feature requests, bug fixes, or pull requests. Feedback and feature requests can be done by opening an issue on the \href{https://github.com/unitaryfund/mitiq}{Mitiq GitHub repository}.
Bug fixes and other pull requests can be done by forking the Mitiq source code, making changes, then opening a pull request to the Mitiq GitHub repository. Pull requests are peer-reviewed by core developers to provide feedback and/or request changes.
Contributors are expected to uphold Mitiq development practices including style guidelines and unit tests. More details can be found in the \href{https://github.com/unitaryfund/mitiq/blob/master/CONTRIBUTING.md}{Contribution guidelines documentation}.

\section{Discussion}\label{sec:discussion}

Now that we have described error mitigation techniques in Mitiq and how to use them, we discuss limitations of these techniques as well as the relationship between zero-noise extrapolation, probabilistic error cancellation, and other strategies.

\subsection{Limitations of zero-noise extrapolation}\label{subsec:zne-limitations}

Zero-noise extrapolation~\cite{Temme_2017_PRL,Li_2017_PRX} is a useful error mitigation technique, but it is not without limitations. First and foremost, the zero-noise estimate is extrapolated, meaning that performance guarantees are quite difficult in general. If a reasonable estimate of how increasing the noise affects the observable (e.g., the blue curves in \cref{fig:ibm-rigetti-randomized-benchmark}) is known, then ZNE can produce good zero-noise estimates. This is the case for simple noise models such as depolarizing noise, but noise in actual quantum systems is more complicated and can produce different behavior than expected, e.g.\ \cref{fig:ibm-rigetti-randomized-benchmark}(b). In this case the performance of ZNE is tied to the performance of the underlying hardware. If expectation values do not produce a smooth curve as noise is increased, the zero-noise estimate may be poor and certain inference techniques may fail. In particular, one has to take into account that any initial error in the measured expectation values will propagate to the zero-noise extrapolation value. This fact can significantly amplify the final estimation uncertainty. In practice, this implies that the evaluation of a mitigated expectation value requires more measurement shots with respect to the unmitigated one.

Additionally, zero-noise extrapolation cannot increase the performance of arbitrary circuits. If the circuit is large enough such that the expectation of the observable is almost constant as noise is increased (e.g., if the state is maximally mixed), then extrapolation will of course not help the zero-noise estimate. The regime in which ZNE is applicable thus depends on the performance of the underlying hardware as well as the circuit. A detailed description of when zero-noise extrapolation is effective, and how effective it is, is the subject of ongoing research.

\subsection{Limitations of probabilistic error cancellation}\label{subsec:pec-limitations}

The limitations of probabilistic error cancellation~\cite{Temme_2017_PRL, endo2018practical} are similar to those of other error mitigation methods: more circuit executions are necessary compared to the unmitigated case and the method it is not appropriate in the asymptotic regime of many gates or large noise.
Compared to ZNE, PEC has the important advantage of producing an unbiased estimation. This means that, if the quasi-probability representations of all the gates are known with sufficiently large accuracy, in the limit of many samples, the PEC estimation converges to the ideal expectation value.
Unfortunately, PEC has some practical disadvantages too. The number of samples grows exponentially with respect to the circuit size and to  the amount of noise. Moreover, the full tomography of the noisy gates is typically necessary in order to build the quasi-probability representations for the ideal gates. One should also take into account that tomographic errors in the characterization of the hardware gates can propagate through the PEC process inducing a significant error in the final estimation.

\subsection{Limitations of Clifford data regression}\label{subsec:cdr-limitations}
Clifford data regression~\cite{czarnik2020error, lowe2020unified} has the promising advantage of being a self-tuning technique since the inference model is not assumed {\it a priori} but learned during the training phase. However, this technique presents some limitations as well. The training phase typically introduces a significant overhead (many training circuits must be executed with both quantum and classical hardware). Moreover, the training data is extracted from near-Clifford circuits which may have a different response to the hardware noise compared to the true circuit of interest. It is also worth noting that this technique requires an efficient classical simulator of near-Clifford circuits in addition to a quantum backend.

\subsection{Overview of error mitigation techniques}\label{subsec:overview-error-mitigation}

Zero-noise extrapolation  was first proposed in~\cite{Temme_2017_PRL,Li_2017_PRX} and first demonstrated experimentally in~\cite{Kandala_2019_Nature}. References~\cite{Giurgica_2020_arxiv,Cai_2020_arxiv} have extended the noise scaling and extrapolation techniques. Additionally, these references and this paper show experimental demonstrations of zero-noise extrapolation and how it can improve the results of noisy quantum computations.

The purposeful randomization of gates
is another approach to quantum error mitigation. Specific techniques include compiling the quantum circuit with twirling gates~\cite{wallman2016noise}, expressing noiseless gates in a basis of noisy gates as in
probabilistic error cancellation~\cite{Temme_2017_PRL},
and a hybrid proposal improving the scaling of the technique with circuit depth and other resources~\cite{endo2018practical}.
Such techniques have been investigated experimentally in trapped ions~\cite{Zhang_2020_NatComm} and superconducting qubits~\cite{Sun_2020_arXiv} (implementing gate set tomography).

Subspace expansion refers to another set of error mitigation techniques. In Ref.~\cite{McClean_2017_PRA}, a hybrid quantum-classical hierarchy was introduced, while in Ref.~\cite{Bonet_2018_PRA}, symmetry verification was introduced. It has been demonstrated with a stabilizer-like method~\cite{McArdle_2019_PRL}, exploiting molecular symmetries~\cite{McClean_2020_NatComm}, and with an experiment on a superconducting circuit device~\cite{Sagastizabal_2019_PRA}. Other symmetry-based protocols have since been proposed~\cite{Koczor_2021,Huggins_2021virtual,Cai_2021}.
Other error mitigation techniques include approximating error-correcting codes in quantum channels~\cite{Cafaro_2014_PRA},
and have been extended to improve quantum sensing~\cite{Otten_2018_PRA}, metrology~\cite{Zhou_2020_PRR}, and reduce errors in analog quantum simulation~\cite{Sun_2020_arXiv}.

\subsection{Differences and relations to neighbouring fields}\label{other-fields}

Quantum error mitigation is deeply connected to quantum error correction and quantum optimal control, two fields of study that also aim at reducing the impact of errors in quantum information processing in quantum computers. More generally, quantum error mitigation is also related to the general theory of open quantum systems. While these are fluid boundaries, it can be useful to point out some differences among these more established fields and the emerging niche of quantum error mitigation.

\subsubsection{Quantum error correction}
Quantum error correction creates logical qubits out of multiple error-prone physical qubits. After applying logical operations which correspond to the physical operations we want to perform in our circuit, ancilla qubits are measured to diagnose which (if any) errors occurred. Depending on the outcome of these ``syndrome measurements'', correction operations are performed to remove the errors (if any) that occurred. If the error rate lies below a certain threshold, errors can be actively removed. We can thus say that the goal of error correction is to detect and exactly correct errors, while the goal of error mitigation is to lessen the effect of errors.

The drawback of quantum error correction techniques is that they require a large overhead in terms of additional physical qubits needed to create logical qubits. Current quantum computing devices have been able to demonstrate some components of quantum error correction with a very small number of qubits~\cite{Gong_2019,Schindler_Barreiro_Monz_Nebendahl_Nigg_Chwalla_Hennrich_Blatt_2011}. Indeed, some techniques for quantum error mitigation emerged as more practical quantum error correction solutions~\cite{Knill_2005_Nature}.

\subsubsection{Quantum optimal control}\label{optimal-control}

Optimal control theory encompasses a versatile set of techniques that can be applied to many scenarios in quantum technology~\cite{Brif_2010_NJP}. It is generally based on a feedback loop between an agent and a target system. A key difference between some quantum error mitigation techniques and quantum optimal control is that the former can be implemented in some instances with post-processing techniques, while the latter relies on an active feedback loop. An example of a specific application of optimal control to quantum dynamics that can be seen as a quantum error mitigation technique is dynamical decoupling~\cite{santos2005dynamical, viola2005random, pokharel2018demonstration}. This technique employs fast control pulses to effectively decouple a system from its environment, with techniques pioneered in the nuclear magnetic resonance community~\cite{Viola_1999_PRL}. Quantum optimal control techniques are being integrated into quantum computing software as a means for noise characterization and error mitigation~\cite{Ball_2020_Arxiv}.

\subsubsection{Environment-induced error protection}\label{open-quantum-systems}
More in general, quantum computing devices can be studied in the framework of open quantum systems~\cite{Carmichael_1999_Springer,Carmichael_2007_Springer,Breuer_2007_Oxford}, that is, systems that exchange energy and information with the surrounding environment in controlled and unwanted ways.

Since errors occur for several reasons in quantum computers, the microscopic description at the physical level can vary broadly, depending on the quantum computing platform that is used as well as the computing architecture, and error mitigation strategies can be employed with an awareness of this variability. For example, superconducting-circuit-based quantum computers have chips that are prone to cross-talk noise~\cite{Murali_2020_ACM}, while qubits encoded in trapped ions need to be shuttled with electromagnetic pulses, and solid-state artificial atoms, including quantum dots, are heavily affected by inhomogeneous broadening~\cite{Buluta_2011_RPP}. Considering the physical layer of the actual device~\cite{Silverio_2021_arXiv,Li_2021_arXiv}, as well as modeling and adapting the control pulses, can in practice result in more effective error mitigation strategies.

One approach to reduce the impact of noise and errors is to tailor a larger computational space to protect  the system from exiting the computational basis. This approach has been particularly fruitful in the context of bosonic quantum codes~\cite{Gottesman_2001_PRA,Mirrahimi_2014_NJP,Michael_2016_PRX,Albert_2020_PRX}.

Moreover, autonomous error correction approaches have been recently proposed and experimentally verified~\cite{Gertler_2021}, which exploit the environment to induce error-robust processes. More in general, decoherence-free subspaces have been proposed within the study of Liouvillian dynamics~\cite{Lidar98,Knill00,Kockum_2018, Lieu_2020}.

\section{Conclusion}\label{sec:conclusion}

We have introduced a fully open-source library for quantum error mitigation on near-term
quantum computers. Our library can interface with multiple quantum programming libraries --- in particular Cirq, Qiskit, pyQuil, and Braket --- and arbitrary quantum processors (real or simulated) available to the user. In this paper, we presented experimental and numerical examples demonstrating how error mitigation can enhance the results of a
noisy quantum computation. We then discussed the library in detail, focusing on the
specific modules of Mitiq associated to different error mitigation techniques: zero-noise extrapolation, probabilistic error cancellation and Clifford data regression. After mentioning additional software information including support and contribution guidelines, we discussed how the error mitigation techniques in our library relate to other error mitigation techniques as well as quantum error correction, quantum optimal control, and the theory of open quantum systems.

In future work, we plan to incorporate additional error mitigation techniques into the library and to expand the set of benchmarks to better understand when quantum error mitigation is beneficial. Work can also be done to improve the existing modules, for example by implementing different noise-scaling methods, inference techniques, or new error cancellation protocols. One candidate noise-scaling method is pulse stretching which will be possible when pulse-level access to quantum hardware becomes available through more cloud services~\cite{Alexander_2020_QST}.
A high-level road map for future development which includes more information on these ideas as well as other ideas can be found on the \href{https://github.com/unitaryfund/mitiq/wiki}{Mitiq Wiki}.

\section*{Acknowledgements}\label{sec:acknowledgements}
This material is based upon work supported by the U.S. Department of Energy, Office of Science, Office of Advanced Scientific Computing Research, Accelerated Research in Quantum Computing under Award Number de-sc0020266 and by IBM under Sponsored Research Agreement No.\ W1975810.
RL acknowledges support from a NASA Space Technology Graduate Research Opportunities Award.
M.H.G is supported by ``la Caixa'' Foundation (ID100010434), Grant No.\ LCF/BQ/DI19/11730056.
PC  was supported by the Laboratory Directed Research and Development (LDRD) program of Los Alamos National Laboratory (LANL) under project numbers 20190659PRD4  and 20210116DR.
We thank IBM and Rigetti for providing access to their quantum computers.
The views expressed in this paper are those of the authors and do not reflect those of IBM or Rigetti.

\bibliographystyle{quantum}
\bibliography{mitiq.bib}

\appendix

\section{Executor examples}\label{sec:executors}

For concreteness, in this appendix we include explicit examples of executor functions which were introduced in \cref{sec:main-usage}. As mentioned, an executor always
accepts a quantum program, sometimes accepts other
arguments, and returns an expectation value as a float.

\subsection{Executors based on real hardware}

Our first executor is the one used in creating \cref{fig:ibm-rigetti-randomized-benchmark}(a). This executor runs a two-qubit circuit on an IBMQ quantum processor and returns the probability of the ground state.
\begin{lstlisting}[language=Python, caption={Defining an executor to run on IBMQ and return the probability of the ground state for a two-qubit circuit. Line 2 requires a valid IBMQ account with saved credentials. We assume that the input \inlinecode{circuit} contains terminal measurements on both qubits.}, firstnumber=1]
import qiskit

provider = qiskit.IBMQ.load_account()

def executor(
    circuit: qiskit.QuantumCircuit,
    backend_name: str = "ibmq_santiago",
    shots: int = 1024,
) -> float:
    # Execute the circuit
    job = qiskit.execute(
        experiments=circuit,
        backend=provider.get_backend(backend_name),
        optimization_level=0,
        shots=shots,
    )
    
    # Get the measurement data
    counts = job.result().get_counts()
    
    # Return the observable
    return counts["00"] / shots
\end{lstlisting}

We also include the same executor function as above but this time running on Rigetti Aspen-8 and used in creating \cref{fig:ibm-rigetti-randomized-benchmark}(b). Note that this executor requires additional steps compared to the same executor in Qiskit --- namely the declaration of classical memory and the addition of measurement operations, as Rigetti QCS handles classical memory different than other platforms. Additionally, it is important to note the use of \inlinecode{basic\_compile} from Mitiq which preserves folded gates when mapping to the native gate set of Aspen-8.
\begin{lstlisting}[language=Python, caption={Defining an executor to run on Rigetti Aspen-8 and return the probability of the ground state. Line 3 requires a Rigetti Quantum Cloud Services (QCS)~\cite{Karalekas_2020} account and reservation. We assume that the input \inlinecode{program} has no measurements, resets, or classical memory declarations.}, firstnumber=1]
import pyquil
from mitiq.mitiq_pyquil.compiler import basic_compile

aspen8 = pyquil.get_qc("Aspen-8", as_qvm=False)

def executor(
    program: pyquil.Program,
    active_reset: bool = True,
    shots: int = 1024,
) -> float:
    prog = Program()

    # Force qubits into the ground state
    if active_reset:
        prog += pyquil.gates.RESET()
        
    # Add the original program
    prog += program.copy()
    
    # Get list of qubits used in the program
    qubits = prog.get_qubits()
    
    # Add classical memory declaration
    ro = prog.declare("ro", "BIT", len(qubits))
    
    # Add measurement operations
    for idx, q in enumerate(qubits):
        prog += MEASURE(q, ro[idx])
    
    # Add number of shots
    prog.wrap_in_numshots_loop(shots)
    
    # Compile the program, keeping folded gates
    prog = basic_compile(prog)

    # Convert to an executable and run
    executable = aspen8.compiler.native_quil_to_executable(prog)
    results = aspen8.run(executable)
    
    # Return the observable
    all_zeros = [sum(b) == 0 for b in results]
    return sum(all_zeros) / shots
\end{lstlisting}

In these examples, we see how the executor function abstracts away details about running on a back-end. This abstraction makes Mitiq compatible with multiple quantum processors using the same interface.

\subsection{Executors based on a classical simulator}

The executor function does not have to use a real quantum processor but instead can use a classical simulator.In this case, the executor is also responsible for adding noise to the circuit. The manner in which noise is added depends on the quantum programming library being used. We show below an example of an executor which adds depolarizing noise to a Cirq circuit and uses density matrix simulation. This executor inputs an arbitrary observable defined by a \inlinecode{cirq.PauliString} and returns its expectation value by sampling.

\begin{lstlisting}[language=Python, 
caption={Cirq executor function based on a density matrix simulation with depolarizing noise and sampling. The
observable is defined via $\texttt{cirq.PauliString}$.}, firstnumber=1]
import cirq

dsim = cirq.DensityMatrixSimulator()

def executor(
    circ: Circuit,
    obs: cirq.PauliString,
    noise: float = 0.01,
    shots: int = 1024,
) -> float:
    # Add depolarizing noise to the circuit
    noisy = circ.with_noise(cirq.depolarize(p=noise))

    # Do the sampling
    psum = cirq.PauliSumCollector(
        noisy,
        obs,
        samples_per_term=shots,
    )
    psum.collect(sampler=dsim)

    # Return the expectation value
    return psum.estimated_energy()
\end{lstlisting}
Other noise models can be easily substituted into this executor by changing the channel in Line 13 from \inlinecode{cirq.depolarize} to a different channel, e.g. \inlinecode{cirq.amplitude\_damp}. Executors using classical simulators in other quantum programming frameworks (e.g., Qiskit or pyQuil) can be defined in an analogous way, although each handles noise in different manners.

Finally, we note that executor functions provided to \inlinecode{execute\_with\_zne} must have only a single argument: the quantum program. The examples above include additional arguments, and it is often convenient to write executors this way. To make an executor with multiple arguments a function of one argument, we can use \inlinecode{functools.partial} as shown below.

\begin{lstlisting}[language=Python, 
caption={Converting a multi-argument executor to a single-argument executor to use with \inlinecode{execute\_with\_zne}. The \inlinecode{functools} library is a built-in Python library.}, firstnumber=1]
from functools import partial

def executor(qprogram, arg1, arg2) -> float:
    ...

new_executor = partial(
    executor,
    arg1=arg1value,
    arg2=arg2value,
)
\end{lstlisting}
The \inlinecode{new\_executor} is now a function of a single argument (the quantum program) and can be used directly with \inlinecode{mitiq.zne.execute\_with\_zne} or \inlinecode{mitiq.pec.execute\_with\_pec}.

\end{document}